\documentclass[aip,jmp,amsmath,amssymb,reprint]{revtex4-1}

\usepackage{graphicx}
\usepackage{dcolumn}
\usepackage{bm}
\usepackage{color}

\usepackage{cases}

\begin{document}

\title{Interplay between the local information based behavioral responses and the epidemic spreading in complex networks}

\author{Can Liu}

\affiliation{School of Mathematical Science, Anhui University, Hefei
230601, P. R. China}

\author{Jia-Rong Xie}
\affiliation{Department of Modern Physics, University of Science and
Technology of China, Hefei 230026, China}

\author{Han-Shuang Chen}
\affiliation{School of Physics and Material Science, Anhui University, Hefei 230601, China}
\author{Hai-Feng Zhang}
\email{haifengzhang1978@gmail.com}
\affiliation{School of Mathematical Science, Anhui University, Hefei
230601, P. R. China}
\affiliation{Center of Information Support \&Assurance Technology, Anhui University, Hefei  230601,  China
}
\affiliation{Department of Communication Engineering, North
University of China, Taiyuan, Shan'xi 030051,  China}

\author{Ming Tang}
\email{tangminghuang521@hotmail.com}
\affiliation{ Web Sciences Center, University of Electronic Science
and Technology of China, Chengdu 611731, China}

\date{\today}

\begin{abstract}
The spreading of an infectious disease can trigger human behavior responses to the disease, which in turn plays a crucial role on the spreading of epidemic. In this study, to illustrate the impacts of the human behavioral responses, a new class of individuals, $S^F$, is introduced to the classical susceptible-infected-recovered ($SIR$) model. In the model, $S^F$ state represents that susceptible individuals who take self-initiate protective measures to lower the probability of being infected, and a susceptible individual may go to $S^F$ state with a response rate when contacting an infectious neighbor. Via the percolation method, the theoretical formulas for the epidemic threshold as well as the prevalence of epidemic are derived. Our finding indicates that, with the increasing of the response rate, the epidemic threshold is enhanced and the prevalence of epidemic is reduced. The analytical results are also verified by the numerical simulations. In addition, we demonstrate that, because the mean field method neglects the dynamic correlations, a wrong result based on the mean field method is obtained---the epidemic threshold is not related to the response rate, i.e., the additional $S^F$ state has no impact on the epidemic threshold.
\end{abstract}

\pacs{89.75.-k, 87.23.Ge, 02.50.Le, 05.65.+b}

\maketitle

\begin{quotation}

Human behavioral responses to the spreading of epidemics have been recognized to have great influence on the epidemic dynamics. Therefore, it is very important to incorporate human behaviors into epidemiological models, which could improve
models' utility in reflecting the reality and studying corresponding controlling measures. However,
analytically-grounded approaches to the problem of interacting effect between epidemic dynamics and human behavioral responses are still lacking so far, and what can help us better understand the impacts of human behavioral responses. In this work, by introducing a new $S^F$ state into classical $SIR$ model to mimic the situation that, when susceptible individuals are aware of the risk of infection, they may take self-protective measures to lower the probability of being infected. We then derive theoretical formulas based on the percolation method for the epidemic threshold and the prevalence of epidemic. Our results indicate that the introduction of the new $S^F$ state can significantly enhance the epidemic threshold and reduce the prevalence of epidemic. It is worth mentioning that a wrong result may be obtained when using the traditional mean field method--the epidemic threshold is not altered by the additional $S^F$ state.  The result highlights that if the effects of human behavioral responses are ignored in mathematical modeling, the obtained results cannot really reflect the spreading mechanism of epidemics among human population.
\end{quotation}

\section{Introduction} \label{sec:intro}
Since network models can well describe the spreading of infectious disease among populations, many efforts have been devoted to studying this field~\cite{newman2003structure,newman2010networks}. At first, researchers mainly paid attention to analyze the impact of the network structure on epidemic spreading and the control strategies, for example, how the network topology affects the epidemic threshold and the prevalence of epidemic~\cite{BBPSV:2005,PhysRevLett.105.218701,newman2002spread,PhysRevE.91.042811,pastor2001epidemic,peter}, or how to design an effective immunization strategy to control the outbreaks of epidemics~\cite{pastor2002immunization,cohen2003efficient}. In reality, outbreaks of epidemics can trigger spontaneous behavioral responses of individuals to take preventive measures, which in turn alters the
epidemic dynamics and affects the disease transmission process~\cite{haifeng2013braess,PhysRevE.86.036117,bauch2003group,bauch2004vaccination,wang2012estimating,funk2010modelling,liu2012impact}. Thus, recently, some studies have made attempts to evaluate the impact of the human behaviors on epidemic dynamics. For instance, Funk \emph{et ~al.}~\cite{funk2009spread} studied
the impacts of awareness spread on both epidemic threshold
and prevalence, and they found that, in a well-mixed population, spread of
awareness can reduce the prevalence of epidemic but does not tend to affect
the epidemic threshold, yet the epidemic threshold is altered when considering on social networks; Sahneh \emph{et al.} considered a
Susceptible-Alter-Infected-Susceptible ($SAIS$) model~\cite{sahneh2012existence},
and they found that the way of behavioral response can enhance the
epidemic threshold; Meloni \emph{et~al.} constructed a meta-population
model incorporating several scenarios of self-initiated
behavioral changes into the mobility patterns of individuals, and
they found that such behavioral changes do not alter the epidemic
threshold, but may increase rather than decrease
the prevalence of epidemic~\cite{meloni2011modeling}. Meanwhile, in Ref.~\cite{wu2012impact}, by designing the transmission
rate of epidemic as a function of the local infected density or the global
infected density, Wu \emph{et~al.}
investigated the effect of such a behavioral
response on the epidemic threshold.

One fact is that, the infectious neighbors can infect a susceptible individual, they can also trigger the awareness of the susceptible individual~\cite{sahneh2012existence,zhang2014suppression}. In view of this, in Ref.~\cite{perra2011towards}, Perra \emph{et~al.} introduced a new class of individuals, $S^F$, that represents susceptible people who self-initiate behavioral
changes that lead to a reduction in the transmissibility of the
infectious disease, into the classical $SIR$ model, and they found that such a model ($SS^FIR$) can induce a rich phase space with multiple epidemic
peaks and tipping points. However,  the network structure was not incorporated into these models. As we know, when the model is considered within the network based framework, new theoretical tools should be used and new phenomena may be observed. In view of this, we incorporate the network structure into the $SS^FIR$ model~\cite{perra2011towards} to investigate its spreading dynamics. In the model, when contacting an infectious neighbor, susceptible individuals may be infected (from $S$ state $I$ state) with a transmission rate or a behavioral response may be triggered (from $S$ state to $S^F$ state) with a response rate. We provide a theoretical formula for the epidemic threshold as well as the prevalence of epidemic via the percolation method~\cite{newman2002spread}, our results show that the introduction of $S^F$ class can enhance the epidemic threshold and reduce the prevalence of epidemic. We also demonstrate that a wrong result can be obtained---the introduction of $S^F$ class cannot alter the epidemic threshold when using mean field method to such a model. The reasons are presented in Sec.~\ref{sec:mean-field}.


\section{Descriptions of the model} \label{sec:model}
For the classical $SIR$ model on complex network, where each node on network can be in one of three states: Susceptible ($S$), Infected ($I$) or Recovered ($R$). The transmission rate along each $SI$ link is $\beta$, and an infected node can enter $R$ state with a recovery rate $\mu$. To reflect the fact that, upon observation of infection, susceptible individuals may adopt protective measures to lower their infection risk, a new class, denoted by $S^F$, is introduced into the original $SIR$ model, we use $SS^FIR$ model to denote the modified $SIR$ model in this study. In the model, when an $S$ node contacts an $I$ neighbor, besides the probability of being infected, the $S$ node can go to $S^F$ state with a response rate $\beta_F\geq0$. The transmission rate for the $S^F$ nodes is lower, thus we assume the transmission rate for $S^F$ nodes is $\gamma\beta$, where $0\leq\gamma<1$ is a discount factor.

The $SS^FIR$ model is described by the four following
reactions and the associated rates:

\begin{eqnarray}\label{1}
S+I\xrightarrow{\beta} 2I,
\end{eqnarray}

\begin{eqnarray}\label{2}
S+I\xrightarrow{\beta_F} S^F+I,
\end{eqnarray}

\begin{eqnarray}\label{3}
S^F+I\xrightarrow{\gamma\beta} 2I,
\end{eqnarray}

\begin{eqnarray}\label{4}
I\xrightarrow{\mu} R.
\end{eqnarray}

Note that the $SS^FIR$ model returns to the $SIR$ model once $\beta_F=0$, and the $S^F$ state corresponds to fully vaccinated state when $\gamma=0$.
\section{Theoretical analysis} \label{sec:theory}

In our model, during a sufficiently small time interval $\Delta t$, the transition\emph{rates} of an $SI$ edge becoming an $II$, $S^FI$ and $SR$ edge are $\beta$, $\beta_F$ and $\mu$, respectively. As a result, the \emph{probabilities} of an $SI$ edge becoming an $II$ and $S^FI$ edge are given as $T_1=\frac{\beta}{\beta+\beta_F+\mu}$ and $T_2=\frac{\beta_F}{\beta+\beta_F+\mu}$, respectively. Similarly, since the transition rate of an $S^FI$ edge becoming an $II$ and $S^FR$ during a sufficiently small time interval $\Delta t$ are $\gamma\beta$ and $\mu$, the probability of an $S^FI$ edge becoming an $II$ edge is $T_3=\frac{\gamma\beta}{\gamma\beta+\mu}$~\cite{hebert2013pathogen}.

To analyze our proposed model, we first define``externally infected neighbor'' (EIN) for
any node. For node $i$, if a neighbor $j$ is an EIN means that $j$ is infected by its neighbors other than $i$ ( i.e., $j$ is infected
even though node $i$ is deleted from the networks, which is the basic assumption of the cavity theory in statistical physics. Note that this method is suitable for the networks with negligible number of loops as the network size is sufficiently large~\cite{newman2013interacting}). The probability of neighbor $j$ being an EIN of $i$
is defined as $\theta$, then, for the a node with degree $k$, the probability of having $m$ EINs is given as:

\begin{eqnarray}\label{5}
p(m|k)=\big(^k_m\big)\theta^m(1-\theta)^{k-m}
\end{eqnarray}

Let $p(R|m)$ be the probability of $i$ being infected when it has number $m$ of EIN. To calculate such a probability, we need to recognize that, in our model, an $S$ node can become an $I$ node through two ways: (a) the $S$ node is directly infected; or (b) the $S$ node first goes to $S^F$ state and then is infected. To facilitate the analysis, we approximately assume that the impacts of $i$'s infected neighbors on node $i$ happen in a non-overlapping order, i.e., they play roles on node $i$ one by one.

For case (a), the probability of node $i$ being infected by the $s$th infected neighbor is given as:
\begin{eqnarray}\label{6}
A_1=(1-T_1-T_2)^{s-1}T_1,
\end{eqnarray}
Eq.~(\ref{6}) indicates that previous $s-1$ infected neighbors have not changed the state of node $i$ (not become $I$ or $S^F$ state) before they become $R$ state. Therefore, the probability of $i$ being infected is:
\begin{eqnarray}\label{7}
p_1(R|m)=\sum_{s=1}^{m}A_1=\frac{T_1}{T_1+T_2}[1-(1-T_1-T_2)^m].
\end{eqnarray}

For case (b), node $i$ should first become $S^F$ state, and the probability that the susceptible node $i$ is altered by the $l$th infected neighbors and becomes $S^F$ can be written as:
\begin{eqnarray}\label{8}
A_2=(1-T_1-T_2)^{l-1}T_2,
\end{eqnarray}
which also indicates that previous $l-1$ infected neighbors have not changed the state of node $i$ before they become $R$ state.
For the remainder $m-l+1$ infected neighbors (including the infected neighbor who just made $i$ go to $S^F$ state), they can infect $i$ with probability:
\begin{eqnarray}\label{9}
A_3=1-(1-T_3)^{m-l+1}.
\end{eqnarray}
As a result, the probability of node $i$ first becoming $S^F$ state and then going to $I$ state is:
\begin{eqnarray}\label{10}
 \nonumber p_2(R|m)&&=\sum\limits_{l= 1}^m {A_2* A_3}
=\frac{{{T_2}}}{{{T_1} + {T_2}}}[1 - {(1 - {T_1} - {T_2})^m}]\\
-&&\frac{{(1 - {T_3}){T_2}}}{{{T_1} + {T_2} - {T_3}}}[{(1 - {T_3})^m} - {(1 - {T_1} - {T_2})^m}].
\end{eqnarray}

The probability $p(R|m)$ is
\begin{eqnarray}\label{11}
\nonumber p(R|m)
&=&p_1(R|m)+p_2(R|m)= 1 - {(1 - {T_1} - {T_2})^m} \\
\nonumber&&-\frac{{(1 - {T_3}){T_2}}}{{{T_1} + {T_2} - {T_3}}}[{(1 - {T_3})^m} - {(1 - {T_1} - {T_2})^m}]\\
\nonumber&=&1 - \frac{{{T_1} - {T_3} + {T_2}{T_3}}}{{{T_1} + {T_2} - {T_3}}}{(1 - {T_1} - {T_2})^m}\\
&& -\frac{{(1 - {T_3}){T_2}}}{{{T_1} + {T_2} - {T_3}}}{(1 - {T_3})^m}.
\end{eqnarray}

Combing Eqs. (\ref{5}) and (\ref{11}), the probability of a node with degree $k$ being infected is
\begin{eqnarray}\label{12}
\nonumber  p(R|k)& =& \sum\limits_m {p(R|m)} p(m|k) = 1 - {(1 - \theta {T_1} - \theta {T_2})^k} \\
\nonumber   && - {{(1 - {T_3}){T_2}} \over {{T_1} + {T_2} - {T_3}}}[{(1 - \theta {T_3})^k} - {(1 - \theta {T_1} - \theta {T_2})^k}]  \\
\nonumber   & =& 1 - {{{T_1} - {T_3} + {T_2}{T_3}} \over {{T_1} + {T_2} - {T_3}}}{(1 - \theta {T_1} - \theta {T_2})^k}\\
&& - {{(1 - {T_3}){T_2}} \over {{T_1} + {T_2} - {T_3}}}{(1 - \theta {T_3})^k}.
 \end{eqnarray}
Then the EIN probability $\theta$ is the solution to the self-consistent condition
\begin{eqnarray}\label{13}
 \nonumber\theta & =& \sum\limits_k {Q(k)p(R|k)}\\
 \nonumber & =& 1 - \frac{{{T_1} - {T_3} + {T_2}{T_3}}}{{{T_1} + {T_2} - {T_3}}}{G_1}(1 - \theta {T_1} - \theta {T_2})\\
&&- \frac{{(1 - {T_3}){T_2}}}{{{T_1} + {T_2} - {T_3}}}{G_1}(1 - \theta {T_3})=f(\theta).
 \end{eqnarray}
In Eq.~(\ref{13}), $Q(k)=\frac{(k+1)P(k+1)}{\langle k \rangle}$ is the excess degree distribution, where $P(k)$ is the degree distribution and $\langle k \rangle$ is the average degree. The generating function for $Q(k)$ is given as
\begin{eqnarray}\label{15}
G_1(x)=\sum\limits_{k=0} {Q(k) x^k}.
\end{eqnarray}

There is a trivial solution $\theta=0$ in self-consistent equation (\ref{13}). To have a non-trivial solution, the following condition must be met:
\begin{equation} \label{16}
\frac{df(\theta)}{d\theta}|_{\theta=0}=(T_1+T_2T_3)
G'_1(1)\geq 1,
\end{equation}
which implies the epidemic can outbreak when
\begin{equation} \label{17}
\frac{\beta}{\mu+\gamma\beta}\frac{\mu+\gamma\beta+\gamma\beta_F}{\mu+\beta+\beta_F}
\geq \frac{\langle k \rangle}{\langle k^2 \rangle-\langle k \rangle}.
\end{equation}

For the prevalence of epidemic (defined as $R(\infty)$), we can numerically solve $\theta$  from self-consistent equation~(\ref{13}), then the formula of $R(\infty)$ is
\begin{eqnarray}\label{18}
 \nonumber R(\infty) & =& \sum\limits_k {P(k)p(R|k)}\\
 \nonumber & =& 1 - \frac{{{T_1} - {T_3} + {T_2}{T_3}}}{{{T_1} + {T_2} - {T_3}}}{G_0}(1 - \theta {T_1} - \theta {T_2})\\
&&- \frac{{(1 - {T_3}){T_2}}}{{{T_1} + {T_2} - {T_3}}}{G_0}(1 - \theta {T_3}).
 \end{eqnarray}

In Eq.~(\ref{18}), $G_0(.)$ is the generating function of degree distribution $P(k)$, which is described as:
\begin{eqnarray}\label{14}
G_0(x)=\sum\limits_{k=0} {P(k) x^k}.
\end{eqnarray}

\section{Simulation results} \label{sec:result}
In this section, we perform an extensive set of Monte Carlo simulations to validate the theoretical predictions in Section~\ref{sec:theory}. Here we carry out simulations on an Erd\H{o}s-R\'{e}nyi network (labeled ER network)~\cite{erdos1960evolution} with network size $N=10000$ and average degree $\langle k \rangle=10$, and a configuration network generated by an uncorrelated configuration model (UCM)~\cite{newman2001random}. The configuration network also has $N=10000$ nodes and the degree
distribution meets $P(k)\sim k^{-3.0}$, whose minimal and maximal degrees are $k_{min}=3$ and
$k_{max}=\sqrt{N}$, respectively.

\subsection{Results on ER network}
Differing from the $SIS$ (Susceptible-Infected-Susceptible) model, it is not an easy thing to determine the epidemic threshold for the $SIR$ model owing to the non-zero value of $R$. In doing so, in Ref.~\cite{shu2014simulated}, Shu et~al. suggested that the variability measure
\begin{eqnarray}\label{measure}
\Delta=\frac{\sqrt{\langle \rho^2\rangle-\langle \rho\rangle^2}}{\langle \rho\rangle}
\end{eqnarray}
can well predict the epidemic threshold for the $SIR$ model, where $\rho$ denotes the prevalence of epidemic in one simulation realization~\cite{crepey2006epidemic,shu2012effects}. $\Delta$ can be explained as the standard deviation of the epidemic prevalence, and is a standard measure to determine critical point in equilibrium phase on magnetic system~\cite{ferreira2011quasistationary}. In our simulations, we have taken at least 1000 independent realizations to predict the epidemic threshold. For convenience, in this study, we set recovery rate $\mu=1.0$.

\begin{figure}
\begin{center}
\includegraphics[width=3.5in]{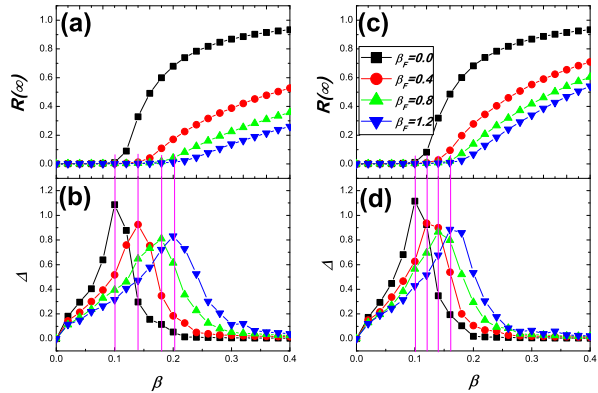}
\caption{(Color online)  On ER networks, the prevalence of epidemic $R(\infty)$ (upper panels) and the variability measure $\Delta$ (lower panels) as functions of the transmission rate $\beta$ for different values of $\beta_F$ and $\gamma$. Note that, here $\beta_F$ is a rate rather probability, as a result, whose value may be larger than one. (a)-(b) $\gamma=0.1$; (c)-(d) $\gamma=0.3$. The pink lines are given to demonstrate that the peak value of $\Delta$ corresponds to the epidemic threshold $\beta_c$.}
\label{fig1}
\end{center}
\end{figure}

In Fig.~\ref{fig1}, for different response rate $\beta_F$, the value of $R(\infty)$ (upper panels) and the measure $\Delta$ (lower panels) as the functions of the transmission rate $\beta$ are investigated. As shown in Fig.~\ref{fig1}, one can observe that the variability measure can well predict the epidemic threshold for our $SS^FIR$ model. As a result, in the following figures, we use this method to determine the epidemic threshold, i.e., the point where the value of  $\Delta$ is the maximal. Fig.~\ref{fig1} also describes that, no matter $\gamma=0.1$ [see Figs.~\ref{fig1}(a)-(b)] or $\gamma=0.3$ [see Figs.~\ref{fig1}(c)-(d)], on the one hand, the epidemic threshold is enhanced as the response rate $\beta_F$ is increased. On the other hand, for the a fixed value of $\beta$, Figs.~\ref{fig1}(a)and (c) suggest that the prevalence of epidemic is remarkably reduced when $\beta_F$ is increased. The results suggest that, by introducing an additional protective state, $S^F$, to the classical $SIR$ model, the conclusions are quite different from the previous results which have not incorporated the impacts of human behavioral responses. The result again emphasizes the fact that the spontaneous behavioral responses of individuals to the emergent diseases have vital impacts on the epidemic dynamics. If the behavioral responses are ignored in mathematical modelling, the obtained results cannot really reflect the spreading mechanism of epidemics among human population.


\begin{figure}
\begin{center}
\includegraphics[width=3.5in]{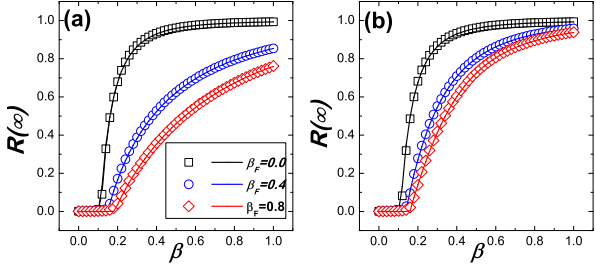}
\caption{(Color online) Comparison between the Monte Carlo based simulations and the theoretical predictions for $R(\infty)$ on ER networks. The simulation results are denoted by symbols and the theoretical predictions are denoted by the corresponding lines. The theoretical results are obtained by substituting Eq.~(\ref{19}) into  Eqs.~(\ref{13}) and (\ref{18}), and the value of $\theta$ is numerically solved from  Eq.~(\ref{13}), then $R(\infty)$ is got from  Eq.~(\ref{18}). (a) $\gamma=0.1$; (b) $\gamma=0.3$. }
\label{fig3}
\end{center}
\end{figure}

\begin{figure}
\begin{center}
\includegraphics[width=3.5in]{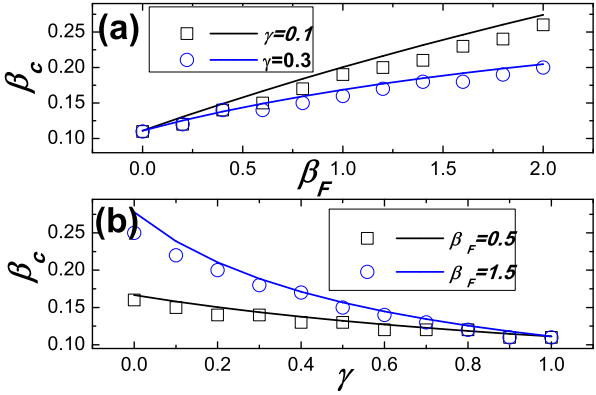}
\caption{(Color online) Comparison between the Monte Carlo based simulations and the theoretical predictions for the epidemic threshold $\beta_c$ on ER networks. The theoretical predictions denoted by lines are obtained from Eq.~(\ref{20}), and the simulation results denoted by symbols are the points where the values $\Delta$ are maximal. (a) $\beta_c$ as a function of the response rate $\beta_F$ for different values of $\gamma$; (b) $\beta_c$ as a function of the discount factor $\gamma$ for different values of $\beta_F$.}
\label{fig4}
\end{center}
\end{figure}

%
We then compare the theoretical results with the Monte Carlo simulations on ER network in Fig.~\ref{fig3} and Fig.~\ref{fig4}. Since the degree distribution of an Erd\H{o}s-R\'{e}nyi network is
${P(k)} = {e^{ -  \langle k \rangle}}\frac{\langle k \rangle^k}{{k!}}$, the generating functions meet:
\begin{equation} \label{19}
G_0(x)=G_1(x)=e^{\langle k \rangle(x-1)}
\end{equation}
According to Ineq.~(\ref{16}), the epidemic threshold $\beta_c$ for ER network is determined by the following equation
\begin{equation} \label{20}
\frac{\beta_c}{\mu+\gamma\beta_c}\frac{\mu+\gamma\beta_c+\gamma\beta_F}{\mu+\beta_c+\beta_F}
= \frac{1}{\langle k \rangle}.
\end{equation}
Moreover, substituting Eq.~(\ref{19}) into Eq.~(\ref{13}) and Eq.~(\ref{18}), the prevalence of epidemic $R(\infty)$ can be easily solved.

The comparison of $R(\infty)$ between the simulations and the theoretical result is plotted in Fig.~\ref{fig3}, which indicates that the numerical simulation and the theoretical result are in good agreement. Meanwhile, the epidemic threshold for $\beta_c$ obtained from Eq.~(\ref{20}) and from numerical method (i.e., the point where $\Delta$ is maximal) is compared in Fig.~\ref{fig4}, which also indicates that the epidemic threshold predicated by our method is remarkable agreement with numerical simulations.  The result in Fig.~\ref{fig4} also suggests that the epidemic threshold $\beta_c$ is increased as the value of $\gamma$ is decreased. Importantly, the reduction is more efficient when the response rate $\beta_F$ is larger.

\subsection{Results on UCM network}

Real complex networked systems often possess certain degree of skewness in
their degree distributions, typically represented by some scale-free topology.
We thus check our model on UCM network with degree distribution $P(k)\sim k^{-3.0}$ to illustrate that our theory can generalize to the networks with heterogenous degree distribution and in the absence of degree-to-degree correlation.

\begin{figure}
\begin{center}
\includegraphics[width=3.5in]{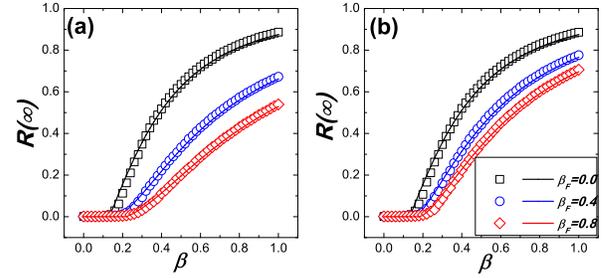}
\caption{(Color online) Comparison between the Monte Carlo based simulations and the theoretical predictions for $R(\infty)$ on UCM networks. The simulation results are denoted by symbols and the theoretical predictions are denoted by the corresponding lines. The theoretical results are obtained by substituting a fixed degree distribution $P(k)$ into  Eqs.~(\ref{14}) and (\ref{14}), and then $R(\infty)$ can be solved from Eqs.~(\ref{14}) and (\ref{14}) as described in Fig.~\ref{fig3}. (a) $\gamma=0.1$; (b) $\gamma=0.3$. }
\label{fig5}
\end{center}
\end{figure}
\begin{figure}
\begin{center}
\includegraphics[width=3.5in]{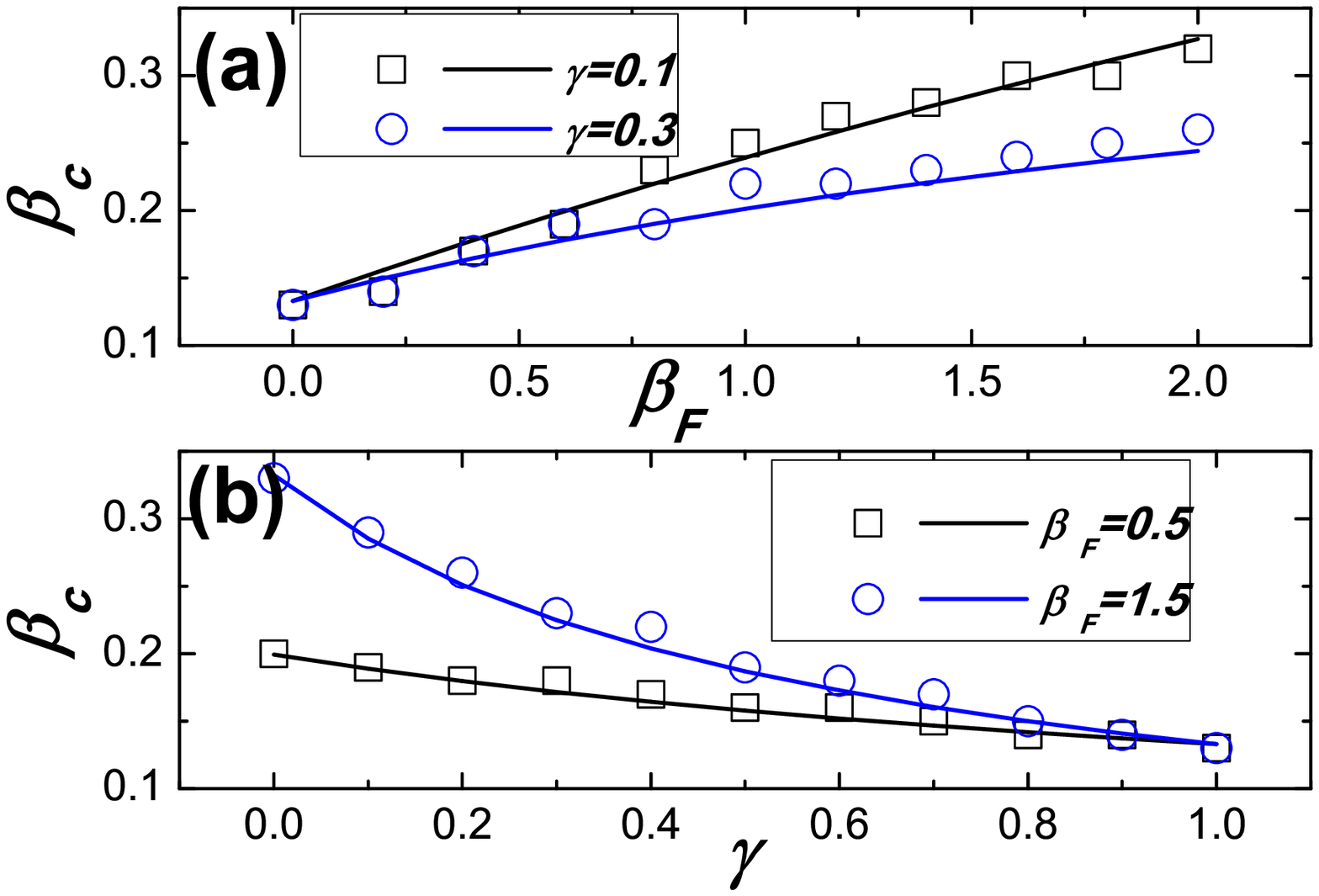}
\caption{(Color online)  Comparison between the Monte Carlo based simulations and the theoretical predictions for the epidemic threshold $\beta_c$ on UCM networks. The theoretical predictions denoted by lines are obtained from Ineq.~(\ref{17}), and the simulation results denoted by symbols are the points where the values $\Delta$ are maximal. (a) $\beta_c$ as a function of the response rate $\beta_F$ for different values of $\gamma$; (b) $\beta_c$ as a function of the discount factor $\gamma$ for different values of $\beta_F$.}
\label{fig6}
\end{center}
\end{figure}

As shown in Fig.~\ref{fig5} and Fig.~\ref{fig6}, one can see that the analytical results are in good agreement with the numerics. They also indicate that, since increasing $\beta_F$ can induce more susceptible individuals go to $S^F$ state and reducing $\gamma$ can lower the risk of $S^F$ nodes, as a result, both of them can lower the prevalence of epidemic $R(\infty)$ and increase the epidemic threshold $\beta_c$.

\begin{figure}
\begin{center}
\includegraphics[width=3.5in]{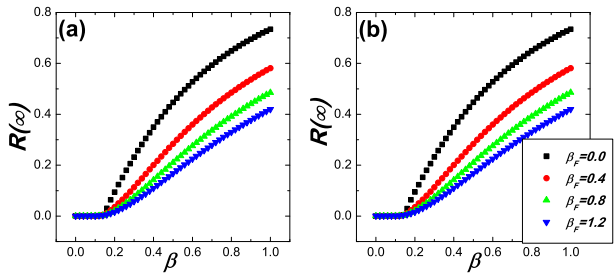}
\caption{(Color online) Based on the differential Eqs.~(\ref{21})-(\ref{22}), the prevalence of epidemic $R(\infty)$ as a function of $\beta$ is presented. (a) $\gamma=0.1$; (b) $\gamma=0.3$. }
\label{fig7}
\end{center}
\end{figure}

\section{Mean field method for the model} \label{sec:mean-field}

One possible way to describe our proposed $SS^{F}IR$ model is the mean field method, which can be written as:
\begin{eqnarray}
 \frac{dS_{k}(t)}{dt}&=&-\beta kS_{k}\Theta-\beta_{F}kS_{k}\Theta,\label{21}\\
 \frac{dS_{k}^{F}(t)}{dt}&=&\beta_{F}kS_{k}\Theta-\beta\gamma kS_{k}^{F}\Theta,\label{22}\\
\frac{dI_{k}(t)}{dt}&=&k\Theta\beta(S_{k}+ \gamma S_{k}^{F})-I_{k},\label{23}\\
\frac{dR_{k}(t)}{dt}&=&I_{k},\label{24}
\end{eqnarray}
where the factor $\Theta(t)=\sum_{k'}P(k'|k)I_{k'}(t)$  represents the
probability that any given link points to an infected node. In absence
of any degree correlations, $\Theta(t)=\frac{1}{\langle k\rangle}\sum_{k}kP(k)I_{k}(t)$~\cite{moreno2002epidemic}.

Based on the above differential equations, we can obtain that the epidemic threshold $\beta_c=\frac{\langle k\rangle}{\langle k^2\rangle}$ (detailed derivation is given in section~\ref{sec:appendix}), which means that the epidemic threshold for our model is not related to the response rate $\beta_F$ or the discount factor $\gamma$, and  which is the same to the epidemic threshold of classical $SIR$ model. The simulation results based on the Eqs.~(\ref{21}-\ref{24}) in Fig.~\ref{fig7} also indicate that, based on mean field method, the epidemic threshold is not altered by different values of $\beta_F$ or $\gamma$. However, our previous simulation results based on Monte Carlo method have suggested that the epidemic threshold $\beta_c$ is increased when $\beta_F$ is increased or $\gamma$ is reduced. That is to say, the conclusion obtained by mean-field method is wrong.

Now let us explain why the mean field method gives a wrong result. It is known that, near the epidemic threshold, the fraction of infected nodes (label $\rho_I$) is very small. When using the mean field method, the dynamic correlation is neglected, the probability of $S$ node becoming $S^F$ is proportional to $O(\rho_I)$ since the average fraction of infected nodes among neighborhood equals to $O(\rho_I)$. Similarly, the probability of $S^F$ node becoming $I$ node is also proportional to $O(\rho_I)$. As a result, the probability of $S\rightarrow S^F\rightarrow I$ is proportional to $O(\rho_I^2)$, which leads to the effect of the $S^F$ is ignored and the epidemic threshold obtained by the mean field method is not related to the value of $\beta_F$ or $\gamma$. In fact, when an $S$ node becomes an $S^F$ node there must exist at least one infected nodes among the neighborhood of the $S$ node. More importantly, these infected neighbors may exist for a certain time interval, so the probability of $S\rightarrow S^F\rightarrow I$ is not proportional to $O(\rho_I^2)$. However, as deduced in Eq.~(\ref{9}), the dynamics correlation near the epidemic threshold is considered in our above analysis, which can accurately predict the epidemic threshold.
\section{Conclusions} \label{sec:conclusion}
To summarize, we have proposed an $SS^FIR$ epidemiological model in complex networks, in which the probability of susceptible individuals becoming $S^F$ state is proportional to the number of infected neighbors, to reflect the fact that individuals are more likely to take protective measures when they find their neighbors are infected. By using theoretical analysis as well as numerical simulations, we found that the prevalence of epidemic is effectively reduced and the epidemic threshold is remarkably increased when the response rate $\beta_F$ is increased or the discount factor $\gamma$ is reduced. Moreover, we have demonstrated that the mean field based analysis provides a wrong result: the epidemic threshold is not related to the response rate $\beta_F$ or discount factor $\gamma$. The reason is that, near the epidemic threshold, the probability of $S\rightarrow S^F\rightarrow I$ is a second order infinitesimal since the mean field method ignores the dynamic correlation, which makes the effect of $S^F$ state to be ignored.

With the development of technology, information induced awareness or self-protective behaviors can not only diffuse through the contact networks where the diseases spread but also
can fast diffuse through many different channels, such as, the word
of mouth, news media, online social networks, and so on. In view of
this, recent well-studied multiplex network theory may be an ideal framework
to mimic the interplay of information or related awareness and the
epidemic dynamics~\cite{granell2013dynamical,wang2014asymmetrically,boccaletti2014structure}. Thus, how to generalize our model to multiplex networks and provide theoretical analysis to the model is a challenge in our further work.
\section{Appendix} \label{sec:appendix}

When assuming $S_{k}(0)\approx1$, then from Eq.~(\ref{21}), we have~\cite{moreno2002epidemic}
\begin{eqnarray}\label{S1}
S_{k}(t)=e^{-k(\beta+\beta^{F})\phi(t)},
\end{eqnarray}
where $\phi(t)=\int_0^{t}\Theta(t')dt'=\frac{1}{\langle k\rangle}\sum_{k}{kP(k)R_k(t)}$.

Substituting Eq.~(\ref{S1}) into Eq.~(\ref{22}), one has:
\begin{eqnarray}\label{S2}
\frac{dS_{k}^{F}}{dt}=\beta_{F}k\Theta(t) e^{-k(\beta+\beta_{F})\phi(t)}-\beta \gamma kS_{k}^{F}\Theta(t).
\end{eqnarray}
By using the variation of constants method, $S_{k}^{F}(t)$ is solved as:
\begin{eqnarray}
\nonumber S_{k}^{F}(t)=\frac{\beta_{F}}{\beta+\beta_{F}-\beta \gamma}e^{-\beta \gamma k\phi (t)}(1-e^{-k(\beta+\beta_{F}-\beta \gamma)\phi(t)}).\\
\end{eqnarray}
Then,
\begin{widetext}
\begin{eqnarray}  \label{S3}
\nonumber \Theta (t)&=&\frac{d\phi(t)}{dt}
=1-\phi(t)-\frac{1}{\langle k\rangle}\Sigma kP(k)e^{-k(\beta+\beta_{F})\phi(t)}\\&&-\frac
{1}{\langle k\rangle}\Sigma kP(k)\frac{\beta_{F}}{\beta+\beta_{F}-\beta \gamma}e^{-\beta \gamma k\phi(t)}(1-e^{-k(\beta+\beta_{F}-\beta \gamma)\phi(t)}).
 \end{eqnarray}
\end{widetext}
 By letting $\phi_{\infty}=\lim_{t\rightarrow\infty}\phi(t)$, and with the fact that $\lim_{t\rightarrow\infty}\frac{d\phi(t)}{dt}=0$ and $I(t\rightarrow\infty)=0$ when the epidemics end, a self-consistent equation can be got from Eq.~(\ref{S3}):

\begin{widetext}
\begin{eqnarray}  \label{S4}
\phi_\infty&=&1-\frac{1}{\langle k\rangle}\Sigma kP(k)e^{-k(\beta+\beta_{F})\phi_\infty}\nonumber\\
&-&\frac
{1}{\langle k\rangle}\Sigma kP(k)\frac{\beta_{F}}{\beta+\beta_{F}-\beta \gamma}e^{-\beta \gamma k\phi_\infty}(1-e^{-k(\beta+\beta_{F}-\beta \gamma)\phi_\infty})=g(\phi_\infty).
 \end{eqnarray}
\end{widetext}
The value $\phi_{\infty}=0$ is always a solution. In order to have a
non-zero solution, the condition
\begin{eqnarray} \label{S5}
\nonumber\frac{df(\phi_\infty)}{d\phi_\infty}|_{\phi_\infty=0}&=&\frac{(\beta+\beta_{F})\langle k^2\rangle}{\langle k\rangle}-\frac{\beta_{F}\langle k^2\rangle}{\langle k\rangle}\\
&=&\frac{\beta\langle k^2\rangle}{\langle k\rangle}\geq 1
\end{eqnarray}
should be satisfied, which means that the epidemic threshold $\beta_c=\frac{\langle k\rangle}{\langle k^2\rangle}$.

\section*{Acknowledgments}

This work is funded by the National Natural Science Foundation of China
(Grant Nos. 61473001, 91324002, 11331009).

\begin{thebibliography}{35}%
\makeatletter
\providecommand \@ifxundefined [1]{%
 \@ifx{#1\undefined}
}%
\providecommand \@ifnum [1]{%
 \ifnum #1\expandafter \@firstoftwo
 \else \expandafter \@secondoftwo
 \fi
}%
\providecommand \@ifx [1]{%
 \ifx #1\expandafter \@firstoftwo
 \else \expandafter \@secondoftwo
 \fi
}%
\providecommand \natexlab [1]{#1}%
\providecommand \enquote  [1]{``#1''}%
\providecommand \bibnamefont  [1]{#1}%
\providecommand \bibfnamefont [1]{#1}%
\providecommand \citenamefont [1]{#1}%
\providecommand \href@noop [0]{\@secondoftwo}%
\providecommand \href [0]{\begingroup \@sanitize@url \@href}%
\providecommand \@href[1]{\@@startlink{#1}\@@href}%
\providecommand \@@href[1]{\endgroup#1\@@endlink}%
\providecommand \@sanitize@url [0]{\catcode `\\12\catcode `\$12\catcode
  `\&12\catcode `\#12\catcode `\^12\catcode `\_12\catcode `\%12\relax}%
\providecommand \@@startlink[1]{}%
\providecommand \@@endlink[0]{}%
\providecommand \url  [0]{\begingroup\@sanitize@url \@url }%
\providecommand \@url [1]{\endgroup\@href {#1}{\urlprefix }}%
\providecommand \urlprefix  [0]{URL }%
\providecommand \Eprint [0]{\href }%
\providecommand \doibase [0]{http://dx.doi.org/}%
\providecommand \selectlanguage [0]{\@gobble}%
\providecommand \bibinfo  [0]{\@secondoftwo}%
\providecommand \bibfield  [0]{\@secondoftwo}%
\providecommand \translation [1]{[#1]}%
\providecommand \BibitemOpen [0]{}%
\providecommand \bibitemStop [0]{}%
\providecommand \bibitemNoStop [0]{.\EOS\space}%
\providecommand \EOS [0]{\spacefactor3000\relax}%
\providecommand \BibitemShut  [1]{\csname bibitem#1\endcsname}%
\let\auto@bib@innerbib\@empty
\bibitem [{\citenamefont {Newman}(2003)}]{newman2003structure}%
  \BibitemOpen
  \bibfield  {author} {\bibinfo {author} {\bibfnamefont {M.~E.~J.}\
  \bibnamefont {Newman}},\ }\href@noop {} {\bibfield  {journal} {\bibinfo
  {journal} {SIAM review}\ }\textbf {\bibinfo {volume} {45}},\ \bibinfo {pages}
  {167} (\bibinfo {year} {2003})}\BibitemShut {NoStop}%
\bibitem [{\citenamefont {Newman}(2010)}]{newman2010networks}%
  \BibitemOpen
  \bibfield  {author} {\bibinfo {author} {\bibfnamefont {M.~E.~J.}\
  \bibnamefont {Newman}},\ }\href@noop {} {\emph {\bibinfo {title} {Networks:
  an introduction}}}\ (\bibinfo  {publisher} {Oxford University Press},\
  \bibinfo {year} {2010})\BibitemShut {NoStop}%
\bibitem [{\citenamefont {Barth\'{e}lemy}\ \emph {et~al.}(2005)\citenamefont
  {Barth\'{e}lemy}, \citenamefont {Barrat}, \citenamefont {Pastor-Satorras},\
  and\ \citenamefont {Vespignani}}]{BBPSV:2005}%
  \BibitemOpen
  \bibfield  {author} {\bibinfo {author} {\bibfnamefont {M.}~\bibnamefont
  {Barth\'{e}lemy}}, \bibinfo {author} {\bibfnamefont {A.}~\bibnamefont
  {Barrat}}, \bibinfo {author} {\bibfnamefont {R.}~\bibnamefont
  {Pastor-Satorras}}, \ and\ \bibinfo {author} {\bibfnamefont {A.}~\bibnamefont
  {Vespignani}},\ }\href@noop {} {\bibfield  {journal} {\bibinfo  {journal}
  {Joural of Theoretical Biology}\ }\textbf {\bibinfo {volume} {235}},\
  \bibinfo {pages} {275} (\bibinfo {year} {2005})}\BibitemShut {NoStop}%
\bibitem [{\citenamefont {Castellano}\ and\ \citenamefont
  {Pastor-Satorras}(2010)}]{PhysRevLett.105.218701}%
  \BibitemOpen
  \bibfield  {author} {\bibinfo {author} {\bibfnamefont {C.}~\bibnamefont
  {Castellano}}\ and\ \bibinfo {author} {\bibfnamefont {R.}~\bibnamefont
  {Pastor-Satorras}},\ }\href@noop {} {\bibfield  {journal} {\bibinfo
  {journal} {Physical Review Letters}\ }\textbf {\bibinfo {volume} {105}},\
  \bibinfo {pages} {218701} (\bibinfo {year} {2010})}\BibitemShut {NoStop}%
\bibitem [{\citenamefont {Newman}(2002)}]{newman2002spread}%
  \BibitemOpen
  \bibfield  {author} {\bibinfo {author} {\bibfnamefont {M.~E.~J.}\
  \bibnamefont {Newman}},\ }\href@noop {} {\bibfield  {journal} {\bibinfo
  {journal} {Physical Review E}\ }\textbf {\bibinfo {volume} {66}},\ \bibinfo
  {pages} {016128} (\bibinfo {year} {2002})}\BibitemShut {NoStop}%
\bibitem [{\citenamefont {Holme}\ and\ \citenamefont
  {Takaguchi}(2015)}]{PhysRevE.91.042811}%
  \BibitemOpen
  \bibfield  {author} {\bibinfo {author} {\bibfnamefont {P.}~\bibnamefont
  {Holme}}\ and\ \bibinfo {author} {\bibfnamefont {T.}~\bibnamefont
  {Takaguchi}},\ }\href {\doibase 10.1103/PhysRevE.91.042811} {\bibfield
  {journal} {\bibinfo  {journal} {Phys. Rev. E}\ }\textbf {\bibinfo {volume}
  {91}},\ \bibinfo {pages} {042811} (\bibinfo {year} {2015})}\BibitemShut
  {NoStop}%
\bibitem [{\citenamefont {Pastor-Satorras}\ and\ \citenamefont
  {Vespignani}(2001)}]{pastor2001epidemic}%
  \BibitemOpen
  \bibfield  {author} {\bibinfo {author} {\bibfnamefont {R.}~\bibnamefont
  {Pastor-Satorras}}\ and\ \bibinfo {author} {\bibfnamefont {A.}~\bibnamefont
  {Vespignani}},\ }\href@noop {} {\bibfield  {journal} {\bibinfo  {journal}
  {Physical Review Letters}\ }\textbf {\bibinfo {volume} {86}},\ \bibinfo
  {pages} {3200} (\bibinfo {year} {2001})}\BibitemShut {NoStop}%
\bibitem [{\citenamefont {Holme}(2013)}]{peter}%
  \BibitemOpen
  \bibfield  {author} {\bibinfo {author} {\bibfnamefont {P.}~\bibnamefont
  {Holme}},\ }\href@noop {} {\bibfield  {journal} {\bibinfo  {journal} {PLoS
  ONE}\ }\textbf {\bibinfo {volume} {8}},\ \bibinfo {pages} {e84429} (\bibinfo
  {year} {2013})}\BibitemShut {NoStop}%
\bibitem [{\citenamefont {Pastor-Satorras}\ and\ \citenamefont
  {Vespignani}(2002)}]{pastor2002immunization}%
  \BibitemOpen
  \bibfield  {author} {\bibinfo {author} {\bibfnamefont {R.}~\bibnamefont
  {Pastor-Satorras}}\ and\ \bibinfo {author} {\bibfnamefont {A.}~\bibnamefont
  {Vespignani}},\ }\href@noop {} {\bibfield  {journal} {\bibinfo  {journal}
  {Physical Review E}\ }\textbf {\bibinfo {volume} {65}},\ \bibinfo {pages}
  {036104} (\bibinfo {year} {2002})}\BibitemShut {NoStop}%
\bibitem [{\citenamefont {Cohen}\ \emph {et~al.}(2003)\citenamefont {Cohen},
  \citenamefont {Havlin},\ and\ \citenamefont
  {Ben-Avraham}}]{cohen2003efficient}%
  \BibitemOpen
  \bibfield  {author} {\bibinfo {author} {\bibfnamefont {R.}~\bibnamefont
  {Cohen}}, \bibinfo {author} {\bibfnamefont {S.}~\bibnamefont {Havlin}}, \
  and\ \bibinfo {author} {\bibfnamefont {D.}~\bibnamefont {Ben-Avraham}},\
  }\href@noop {} {\bibfield  {journal} {\bibinfo  {journal} {Physical Review
  Letters}\ }\textbf {\bibinfo {volume} {91}},\ \bibinfo {pages} {247901}
  (\bibinfo {year} {2003})}\BibitemShut {NoStop}%
\bibitem [{\citenamefont {Zhang}\ \emph {et~al.}(2013)\citenamefont {Zhang},
  \citenamefont {Yang}, \citenamefont {Wu}, \citenamefont {Wang},\ and\
  \citenamefont {Zhou}}]{haifeng2013braess}%
  \BibitemOpen
  \bibfield  {author} {\bibinfo {author} {\bibfnamefont {H.-F.}\ \bibnamefont
  {Zhang}}, \bibinfo {author} {\bibfnamefont {Z.}~\bibnamefont {Yang}},
  \bibinfo {author} {\bibfnamefont {Z.-X.}\ \bibnamefont {Wu}}, \bibinfo
  {author} {\bibfnamefont {B.-H.}\ \bibnamefont {Wang}}, \ and\ \bibinfo
  {author} {\bibfnamefont {T.}~\bibnamefont {Zhou}},\ }\href@noop {} {\bibfield
   {journal} {\bibinfo  {journal} {Scientific Reports}\ }\textbf {\bibinfo
  {volume} {3}},\ \bibinfo {pages} {3292} (\bibinfo {year} {2013})}\BibitemShut
  {NoStop}%
\bibitem [{\citenamefont {Ruan}\ \emph {et~al.}(2012)\citenamefont {Ruan},
  \citenamefont {Tang},\ and\ \citenamefont {Liu}}]{PhysRevE.86.036117}%
  \BibitemOpen
  \bibfield  {author} {\bibinfo {author} {\bibfnamefont {Z.}~\bibnamefont
  {Ruan}}, \bibinfo {author} {\bibfnamefont {M.}~\bibnamefont {Tang}}, \ and\
  \bibinfo {author} {\bibfnamefont {Z.}~\bibnamefont {Liu}},\ }\href {\doibase
  10.1103/PhysRevE.86.036117} {\bibfield  {journal} {\bibinfo  {journal} {Phys.
  Rev. E}\ }\textbf {\bibinfo {volume} {86}},\ \bibinfo {pages} {036117}
  (\bibinfo {year} {2012})}\BibitemShut {NoStop}%
\bibitem [{\citenamefont {Bauch}\ \emph {et~al.}(2003)\citenamefont {Bauch},
  \citenamefont {Galvani},\ and\ \citenamefont {Earn}}]{bauch2003group}%
  \BibitemOpen
  \bibfield  {author} {\bibinfo {author} {\bibfnamefont {C.~T.}\ \bibnamefont
  {Bauch}}, \bibinfo {author} {\bibfnamefont {A.~P.}\ \bibnamefont {Galvani}},
  \ and\ \bibinfo {author} {\bibfnamefont {D.~J.}\ \bibnamefont {Earn}},\
  }\href@noop {} {\bibfield  {journal} {\bibinfo  {journal} {Proceedings of the
  National Academy of Sciences}\ }\textbf {\bibinfo {volume} {100}},\ \bibinfo
  {pages} {10564} (\bibinfo {year} {2003})}\BibitemShut {NoStop}%
\bibitem [{\citenamefont {Bauch}\ and\ \citenamefont
  {Earn}(2004)}]{bauch2004vaccination}%
  \BibitemOpen
  \bibfield  {author} {\bibinfo {author} {\bibfnamefont {C.~T.}\ \bibnamefont
  {Bauch}}\ and\ \bibinfo {author} {\bibfnamefont {D.~J.}\ \bibnamefont
  {Earn}},\ }\href@noop {} {\bibfield  {journal} {\bibinfo  {journal}
  {Proceedings of the National Academy of Sciences of the United States of
  America}\ }\textbf {\bibinfo {volume} {101}},\ \bibinfo {pages} {13391}
  (\bibinfo {year} {2004})}\BibitemShut {NoStop}%
\bibitem [{\citenamefont {Wang}\ \emph {et~al.}(2012)\citenamefont {Wang},
  \citenamefont {Zhang}, \citenamefont {Huang},\ and\ \citenamefont
  {Li}}]{wang2012estimating}%
  \BibitemOpen
  \bibfield  {author} {\bibinfo {author} {\bibfnamefont {L.}~\bibnamefont
  {Wang}}, \bibinfo {author} {\bibfnamefont {Y.}~\bibnamefont {Zhang}},
  \bibinfo {author} {\bibfnamefont {T.}~\bibnamefont {Huang}}, \ and\ \bibinfo
  {author} {\bibfnamefont {X.}~\bibnamefont {Li}},\ }\href@noop {} {\bibfield
  {journal} {\bibinfo  {journal} {Physical Review E}\ }\textbf {\bibinfo
  {volume} {86}},\ \bibinfo {pages} {032901} (\bibinfo {year}
  {2012})}\BibitemShut {NoStop}%
\bibitem [{\citenamefont {Funk}\ \emph {et~al.}(2010)\citenamefont {Funk},
  \citenamefont {Salath{\'e}},\ and\ \citenamefont
  {Jansen}}]{funk2010modelling}%
  \BibitemOpen
  \bibfield  {author} {\bibinfo {author} {\bibfnamefont {S.}~\bibnamefont
  {Funk}}, \bibinfo {author} {\bibfnamefont {M.}~\bibnamefont {Salath{\'e}}}, \
  and\ \bibinfo {author} {\bibfnamefont {V.~A.}\ \bibnamefont {Jansen}},\
  }\href@noop {} {\bibfield  {journal} {\bibinfo  {journal} {Journal of The
  Royal Society Interface}\ }\textbf {\bibinfo {volume} {7}},\ \bibinfo {pages}
  {1247} (\bibinfo {year} {2010})}\BibitemShut {NoStop}%
\bibitem [{\citenamefont {Liu}\ \emph {et~al.}(2012)\citenamefont {Liu},
  \citenamefont {Wu},\ and\ \citenamefont {Zhang}}]{liu2012impact}%
  \BibitemOpen
  \bibfield  {author} {\bibinfo {author} {\bibfnamefont {X.-T.}\ \bibnamefont
  {Liu}}, \bibinfo {author} {\bibfnamefont {Z.-X.}\ \bibnamefont {Wu}}, \ and\
  \bibinfo {author} {\bibfnamefont {L.}~\bibnamefont {Zhang}},\ }\href@noop {}
  {\bibfield  {journal} {\bibinfo  {journal} {Physical Review E}\ }\textbf
  {\bibinfo {volume} {86}},\ \bibinfo {pages} {051132} (\bibinfo {year}
  {2012})}\BibitemShut {NoStop}%
\bibitem [{\citenamefont {Funk}\ \emph {et~al.}(2009)\citenamefont {Funk},
  \citenamefont {Gilad}, \citenamefont {Watkins},\ and\ \citenamefont
  {Jansen}}]{funk2009spread}%
  \BibitemOpen
  \bibfield  {author} {\bibinfo {author} {\bibfnamefont {S.}~\bibnamefont
  {Funk}}, \bibinfo {author} {\bibfnamefont {E.}~\bibnamefont {Gilad}},
  \bibinfo {author} {\bibfnamefont {C.}~\bibnamefont {Watkins}}, \ and\
  \bibinfo {author} {\bibfnamefont {V.~A.}\ \bibnamefont {Jansen}},\
  }\href@noop {} {\bibfield  {journal} {\bibinfo  {journal} {Proceedings of the
  National Academy of Sciences}\ }\textbf {\bibinfo {volume} {106}},\ \bibinfo
  {pages} {6872} (\bibinfo {year} {2009})}\BibitemShut {NoStop}%
\bibitem [{\citenamefont {Sahneh}\ \emph {et~al.}(2012)\citenamefont {Sahneh},
  \citenamefont {Chowdhury},\ and\ \citenamefont
  {Scoglio}}]{sahneh2012existence}%
  \BibitemOpen
  \bibfield  {author} {\bibinfo {author} {\bibfnamefont {F.~D.}\ \bibnamefont
  {Sahneh}}, \bibinfo {author} {\bibfnamefont {F.~N.}\ \bibnamefont
  {Chowdhury}}, \ and\ \bibinfo {author} {\bibfnamefont {C.~M.}\ \bibnamefont
  {Scoglio}},\ }\href@noop {} {\bibfield  {journal} {\bibinfo  {journal}
  {Scientific Reports}\ }\textbf {\bibinfo {volume} {2}},\ \bibinfo {pages}
  {632} (\bibinfo {year} {2012})}\BibitemShut {NoStop}%
\bibitem [{\citenamefont {Meloni}\ \emph {et~al.}(2011)\citenamefont {Meloni},
  \citenamefont {Perra}, \citenamefont {Arenas}, \citenamefont {G{\'o}mez},
  \citenamefont {Moreno},\ and\ \citenamefont
  {Vespignani}}]{meloni2011modeling}%
  \BibitemOpen
  \bibfield  {author} {\bibinfo {author} {\bibfnamefont {S.}~\bibnamefont
  {Meloni}}, \bibinfo {author} {\bibfnamefont {N.}~\bibnamefont {Perra}},
  \bibinfo {author} {\bibfnamefont {A.}~\bibnamefont {Arenas}}, \bibinfo
  {author} {\bibfnamefont {S.}~\bibnamefont {G{\'o}mez}}, \bibinfo {author}
  {\bibfnamefont {Y.}~\bibnamefont {Moreno}}, \ and\ \bibinfo {author}
  {\bibfnamefont {A.}~\bibnamefont {Vespignani}},\ }\href@noop {} {\bibfield
  {journal} {\bibinfo  {journal} {Scientific Reports}\ }\textbf {\bibinfo
  {volume} {1}},\ \bibinfo {pages} {62} (\bibinfo {year} {2011})}\BibitemShut
  {NoStop}%
\bibitem [{\citenamefont {Wu}\ \emph {et~al.}(2012)\citenamefont {Wu},
  \citenamefont {Fu}, \citenamefont {Small},\ and\ \citenamefont
  {Xu}}]{wu2012impact}%
  \BibitemOpen
  \bibfield  {author} {\bibinfo {author} {\bibfnamefont {Q.}~\bibnamefont
  {Wu}}, \bibinfo {author} {\bibfnamefont {X.}~\bibnamefont {Fu}}, \bibinfo
  {author} {\bibfnamefont {M.}~\bibnamefont {Small}}, \ and\ \bibinfo {author}
  {\bibfnamefont {X.-J.}\ \bibnamefont {Xu}},\ }\href@noop {} {\bibfield
  {journal} {\bibinfo  {journal} {Chaos}\ }\textbf {\bibinfo {volume} {22}},\
  \bibinfo {pages} {013101} (\bibinfo {year} {2012})}\BibitemShut {NoStop}%
\bibitem [{\citenamefont {Zhang}\ \emph {et~al.}(2014)\citenamefont {Zhang},
  \citenamefont {Xie}, \citenamefont {Tang},\ and\ \citenamefont
  {Lai}}]{zhang2014suppression}%
  \BibitemOpen
  \bibfield  {author} {\bibinfo {author} {\bibfnamefont {H.-F.}\ \bibnamefont
  {Zhang}}, \bibinfo {author} {\bibfnamefont {J.-R.}\ \bibnamefont {Xie}},
  \bibinfo {author} {\bibfnamefont {M.}~\bibnamefont {Tang}}, \ and\ \bibinfo
  {author} {\bibfnamefont {Y.-C.}\ \bibnamefont {Lai}},\ }\href@noop {}
  {\bibfield  {journal} {\bibinfo  {journal} {Chaos}\ }\textbf {\bibinfo
  {volume} {24}},\ \bibinfo {pages} {043106} (\bibinfo {year}
  {2014})}\BibitemShut {NoStop}%
\bibitem [{\citenamefont {Perra}\ \emph {et~al.}(2011)\citenamefont {Perra},
  \citenamefont {Balcan}, \citenamefont {Gon{\c{c}}alves},\ and\ \citenamefont
  {Vespignani}}]{perra2011towards}%
  \BibitemOpen
  \bibfield  {author} {\bibinfo {author} {\bibfnamefont {N.}~\bibnamefont
  {Perra}}, \bibinfo {author} {\bibfnamefont {D.}~\bibnamefont {Balcan}},
  \bibinfo {author} {\bibfnamefont {B.}~\bibnamefont {Gon{\c{c}}alves}}, \ and\
  \bibinfo {author} {\bibfnamefont {A.}~\bibnamefont {Vespignani}},\
  }\href@noop {} {\bibfield  {journal} {\bibinfo  {journal} {PLoS One}\
  }\textbf {\bibinfo {volume} {6}},\ \bibinfo {pages} {e23084} (\bibinfo {year}
  {2011})}\BibitemShut {NoStop}%
\bibitem [{\citenamefont {H{\'e}bert-Dufresne}\ \emph
  {et~al.}(2013)\citenamefont {H{\'e}bert-Dufresne}, \citenamefont
  {Patterson-Lomba}, \citenamefont {Goerg},\ and\ \citenamefont
  {Althouse}}]{hebert2013pathogen}%
  \BibitemOpen
  \bibfield  {author} {\bibinfo {author} {\bibfnamefont {L.}~\bibnamefont
  {H{\'e}bert-Dufresne}}, \bibinfo {author} {\bibfnamefont {O.}~\bibnamefont
  {Patterson-Lomba}}, \bibinfo {author} {\bibfnamefont {G.~M.}\ \bibnamefont
  {Goerg}}, \ and\ \bibinfo {author} {\bibfnamefont {B.~M.}\ \bibnamefont
  {Althouse}},\ }\href@noop {} {\bibfield  {journal} {\bibinfo  {journal}
  {Physical Review Letters}\ }\textbf {\bibinfo {volume} {110}},\ \bibinfo
  {pages} {108103} (\bibinfo {year} {2013})}\BibitemShut {NoStop}%
\bibitem [{\citenamefont {Newman}\ and\ \citenamefont
  {Ferrario}(2013)}]{newman2013interacting}%
  \BibitemOpen
  \bibfield  {author} {\bibinfo {author} {\bibfnamefont {M.~E.~J.}\
  \bibnamefont {Newman}}\ and\ \bibinfo {author} {\bibfnamefont {C.~R.}\
  \bibnamefont {Ferrario}},\ }\href@noop {} {\bibfield  {journal} {\bibinfo
  {journal} {PLoS ONE}\ }\textbf {\bibinfo {volume} {8}},\ \bibinfo {pages}
  {e71321} (\bibinfo {year} {2013})}\BibitemShut {NoStop}%
\bibitem [{\citenamefont {Erd\H{o}s}\ and\ \citenamefont
  {R{\'e}nyi}(1960)}]{erdos1960evolution}%
  \BibitemOpen
  \bibfield  {author} {\bibinfo {author} {\bibfnamefont {P.}~\bibnamefont
  {Erd\H{o}s}}\ and\ \bibinfo {author} {\bibfnamefont {A.}~\bibnamefont
  {R{\'e}nyi}},\ }\href@noop {} {\bibfield  {journal} {\bibinfo  {journal}
  {Publ. Math. Inst. Hungar. Acad. Sci}\ }\textbf {\bibinfo {volume} {5}},\
  \bibinfo {pages} {17} (\bibinfo {year} {1960})}\BibitemShut {NoStop}%
\bibitem [{\citenamefont {Newman}\ \emph {et~al.}(2001)\citenamefont {Newman},
  \citenamefont {Strogatz},\ and\ \citenamefont {Watts}}]{newman2001random}%
  \BibitemOpen
  \bibfield  {author} {\bibinfo {author} {\bibfnamefont {M.~E.~J.}\
  \bibnamefont {Newman}}, \bibinfo {author} {\bibfnamefont {S.~H.}\
  \bibnamefont {Strogatz}}, \ and\ \bibinfo {author} {\bibfnamefont {D.~J.}\
  \bibnamefont {Watts}},\ }\href@noop {} {\bibfield  {journal} {\bibinfo
  {journal} {Physical Review E}\ }\textbf {\bibinfo {volume} {64}},\ \bibinfo
  {pages} {026118} (\bibinfo {year} {2001})}\BibitemShut {NoStop}%
\bibitem [{\citenamefont {Shu}\ \emph {et~al.}(2014)\citenamefont {Shu},
  \citenamefont {Wang},\ and\ \citenamefont {Tang}}]{shu2014simulated}%
  \BibitemOpen
  \bibfield  {author} {\bibinfo {author} {\bibfnamefont {P.}~\bibnamefont
  {Shu}}, \bibinfo {author} {\bibfnamefont {W.}~\bibnamefont {Wang}}, \ and\
  \bibinfo {author} {\bibfnamefont {M.}~\bibnamefont {Tang}},\ }\href@noop {}
  {\bibfield  {journal} {\bibinfo  {journal} {arXiv preprint arXiv:1410.0459}\
  } (\bibinfo {year} {2014})}\BibitemShut {NoStop}%
\bibitem [{\citenamefont {Crepey}\ \emph {et~al.}(2006)\citenamefont {Crepey},
  \citenamefont {Alvarez},\ and\ \citenamefont
  {Barth{\'e}lemy}}]{crepey2006epidemic}%
  \BibitemOpen
  \bibfield  {author} {\bibinfo {author} {\bibfnamefont {P.}~\bibnamefont
  {Crepey}}, \bibinfo {author} {\bibfnamefont {F.~P.}\ \bibnamefont {Alvarez}},
  \ and\ \bibinfo {author} {\bibfnamefont {M.}~\bibnamefont {Barth{\'e}lemy}},\
  }\href@noop {} {\bibfield  {journal} {\bibinfo  {journal} {Physical Review
  E}\ }\textbf {\bibinfo {volume} {73}},\ \bibinfo {pages} {046131} (\bibinfo
  {year} {2006})}\BibitemShut {NoStop}%
\bibitem [{\citenamefont {Shu}\ \emph {et~al.}(2012)\citenamefont {Shu},
  \citenamefont {Tang}, \citenamefont {Gong},\ and\ \citenamefont
  {Liu}}]{shu2012effects}%
  \BibitemOpen
  \bibfield  {author} {\bibinfo {author} {\bibfnamefont {P.}~\bibnamefont
  {Shu}}, \bibinfo {author} {\bibfnamefont {M.}~\bibnamefont {Tang}}, \bibinfo
  {author} {\bibfnamefont {K.}~\bibnamefont {Gong}}, \ and\ \bibinfo {author}
  {\bibfnamefont {Y.}~\bibnamefont {Liu}},\ }\href@noop {} {\bibfield
  {journal} {\bibinfo  {journal} {Chaos}\ }\textbf {\bibinfo {volume} {22}},\
  \bibinfo {pages} {043124} (\bibinfo {year} {2012})}\BibitemShut {NoStop}%
\bibitem [{\citenamefont {Ferreira}\ \emph {et~al.}(2011)\citenamefont
  {Ferreira}, \citenamefont {Ferreira}, \citenamefont {Castellano},\ and\
  \citenamefont {Pastor-Satorras}}]{ferreira2011quasistationary}%
  \BibitemOpen
  \bibfield  {author} {\bibinfo {author} {\bibfnamefont {S.~C.}\ \bibnamefont
  {Ferreira}}, \bibinfo {author} {\bibfnamefont {R.~S.}\ \bibnamefont
  {Ferreira}}, \bibinfo {author} {\bibfnamefont {C.}~\bibnamefont
  {Castellano}}, \ and\ \bibinfo {author} {\bibfnamefont {R.}~\bibnamefont
  {Pastor-Satorras}},\ }\href@noop {} {\bibfield  {journal} {\bibinfo
  {journal} {Physical Review E}\ }\textbf {\bibinfo {volume} {84}},\ \bibinfo
  {pages} {066102} (\bibinfo {year} {2011})}\BibitemShut {NoStop}%
\bibitem [{\citenamefont {Moreno}\ \emph {et~al.}(2002)\citenamefont {Moreno},
  \citenamefont {Pastor-Satorras},\ and\ \citenamefont
  {Vespignani}}]{moreno2002epidemic}%
  \BibitemOpen
  \bibfield  {author} {\bibinfo {author} {\bibfnamefont {Y.}~\bibnamefont
  {Moreno}}, \bibinfo {author} {\bibfnamefont {R.}~\bibnamefont
  {Pastor-Satorras}}, \ and\ \bibinfo {author} {\bibfnamefont {A.}~\bibnamefont
  {Vespignani}},\ }\href@noop {} {\bibfield  {journal} {\bibinfo  {journal}
  {The European Physical Journal B}\ }\textbf {\bibinfo {volume} {26}},\
  \bibinfo {pages} {521} (\bibinfo {year} {2002})}\BibitemShut {NoStop}%
\bibitem [{\citenamefont {Granell}\ \emph {et~al.}(2013)\citenamefont
  {Granell}, \citenamefont {Gomez},\ and\ \citenamefont
  {Arenas}}]{granell2013dynamical}%
  \BibitemOpen
  \bibfield  {author} {\bibinfo {author} {\bibfnamefont {C.}~\bibnamefont
  {Granell}}, \bibinfo {author} {\bibfnamefont {S.}~\bibnamefont {Gomez}}, \
  and\ \bibinfo {author} {\bibfnamefont {A.}~\bibnamefont {Arenas}},\
  }\href@noop {} {\bibfield  {journal} {\bibinfo  {journal} {Physical Review
  Letters}\ }\textbf {\bibinfo {volume} {111}},\ \bibinfo {pages} {128701}
  (\bibinfo {year} {2013})}\BibitemShut {NoStop}%
\bibitem [{\citenamefont {Wang}\ \emph {et~al.}(2014)\citenamefont {Wang},
  \citenamefont {Tang}, \citenamefont {Yang}, \citenamefont {Do}, \citenamefont
  {Lai},\ and\ \citenamefont {Lee}}]{wang2014asymmetrically}%
  \BibitemOpen
  \bibfield  {author} {\bibinfo {author} {\bibfnamefont {W.}~\bibnamefont
  {Wang}}, \bibinfo {author} {\bibfnamefont {M.}~\bibnamefont {Tang}}, \bibinfo
  {author} {\bibfnamefont {H.}~\bibnamefont {Yang}}, \bibinfo {author}
  {\bibfnamefont {Y.}~\bibnamefont {Do}}, \bibinfo {author} {\bibfnamefont
  {Y.-C.}\ \bibnamefont {Lai}}, \ and\ \bibinfo {author} {\bibfnamefont
  {G.}~\bibnamefont {Lee}},\ }\href@noop {} {\bibfield  {journal} {\bibinfo
  {journal} {Scientific Reports}\ }\textbf {\bibinfo {volume} {4}},\ \bibinfo
  {pages} {5097} (\bibinfo {year} {2014})}\BibitemShut {NoStop}%
\bibitem [{\citenamefont {Boccaletti}\ \emph {et~al.}(2014)\citenamefont
  {Boccaletti}, \citenamefont {Bianconi}, \citenamefont {Criado}, \citenamefont
  {Del~Genio}, \citenamefont {G{\'o}mez-Garde{\~n}es}, \citenamefont {Romance},
  \citenamefont {Sendi{\~n}a-Nadal}, \citenamefont {Wang},\ and\ \citenamefont
  {Zanin}}]{boccaletti2014structure}%
  \BibitemOpen
  \bibfield  {author} {\bibinfo {author} {\bibfnamefont {S.}~\bibnamefont
  {Boccaletti}}, \bibinfo {author} {\bibfnamefont {G.}~\bibnamefont
  {Bianconi}}, \bibinfo {author} {\bibfnamefont {R.}~\bibnamefont {Criado}},
  \bibinfo {author} {\bibfnamefont {C.}~\bibnamefont {Del~Genio}}, \bibinfo
  {author} {\bibfnamefont {J.}~\bibnamefont {G{\'o}mez-Garde{\~n}es}}, \bibinfo
  {author} {\bibfnamefont {M.}~\bibnamefont {Romance}}, \bibinfo {author}
  {\bibfnamefont {I.}~\bibnamefont {Sendi{\~n}a-Nadal}}, \bibinfo {author}
  {\bibfnamefont {Z.}~\bibnamefont {Wang}}, \ and\ \bibinfo {author}
  {\bibfnamefont {M.}~\bibnamefont {Zanin}},\ }\href@noop {} {\bibfield
  {journal} {\bibinfo  {journal} {Physics Reports}\ }\textbf {\bibinfo {volume}
  {544}},\ \bibinfo {pages} {1} (\bibinfo {year} {2014})}\BibitemShut {NoStop}%
\end{thebibliography}
%

\end{document}